\documentclass[onecolumn,12pt,draftcls]{IEEEtran}

\usepackage{./report_default}
\usepackage{soul}

\newcommand{\tilnu}{\tilde{\nu}}

\newcommand{\lp}{\left(}
\newcommand{\rp}{\right)}
\newcommand{\lb}{\left[}
\newcommand{\rb}{\right]}
\newcommand{\lbp}{\left\{}
\newcommand{\rbp}{\right\}}

\newcommand{\mcal}{\mathcal}

\newcommand{\mbb}{\mathbb}

\newcommand{\lfl}{\left\lfloor}
\newcommand{\rfl}{\right\rfloor}

\newcommand{\diid}{\overset{\text{i.i.d.}}{\sim}}

\newcommand{\E}{\mathbb{E}}

\renewcommand{\Pr}{\mathbb{P}}

\makeatletter
\newcommand*{\indep}{%
  \mathbin{%
    \mathpalette{\@indep}{}%
  }%
}
\newcommand*{\nindep}{%
  \mathbin{
    \mathpalette{\@indep}{\not}
  }%
}
\newcommand*{\@indep}[2]{%
  \sbox0{$#1\perp\m@th$}
  \sbox2{$#1=$}
  \sbox4{$#1\vcenter{}$}
  \rlap{\copy0}
  \dimen@=\dimexpr\ht2-\ht4-.2pt\relax
  \kern\dimen@
  {#2}%
  \kern\dimen@
  \copy0 
} 
\makeatother

\newtheorem{theorem}{Theorem}
\newtheorem{corollary}{Corollary}
\newtheorem{lemma}{Lemma}

\newcommand{\vvv}[1]{{\color{red}{#1}}}

\begin{document}

\allowdisplaybreaks

\onecolumn 

\title{High Probability Latency Quickest Change Detection over a Finite Horizon}

\author{\IEEEauthorblockN{Yu-Han Huang\IEEEauthorrefmark{1} and Venugopal V. Veeravalli\IEEEauthorrefmark{5}}
\IEEEauthorblockA{Coordinated Science Lab, University of Illinois Urbana-Champaign \\ Email:  \IEEEauthorrefmark{1}yuhanhh2@illinois.edu}}
\maketitle
\begin{abstract}
    THIS PAPER IS ELIGIBLE FOR THE STUDENT PAPER AWARD. 
A finite horizon variant of the quickest change detection problem is studied, in which the goal is to minimize a delay threshold (latency), under constraints on the probability of false alarm and the probability that the latency is exceeded. In addition, the horizon is not known to the change detector. A variant of the cumulative sum (CuSum) test with a threshold that increasing logarithmically with time is proposed as a candidate solution to the problem. An information-theoretic lower bound on the minimum value of the latency under the constraints is then developed. This
lower bound is used to establish certain asymptotic optimality properties of the proposed test in terms of the horizon
and the false alarm probability. Some experimental results are given to illustrate the performance of the test.

\end{abstract}


\section{Introduction}

\label{sec:Intro}

The problem of detecting changes or anomalies in stochastic systems and time series, often referred to as the quickest change detection (QCD) problem, arises in various engineering and scientific settings. The observations of the system are assumed to undergo a change in distribution at the change-point, and the goal is to detect this change as soon as possible, subject to false alarm constraints. See \cite{poor-hadj-qcd-book-2009,tart-niki-bass-2014,vvv_qcd_overview,xie_vvv_qcd_overview} for books and survey articles on the topic. 


The QCD problem is formulated mathematically as a constrained optimization problem to minimize a measure of detection
delay, subject to a constraint on an appropriate false alarm metric. The most commonly used false alarm metric is the \emph{mean time to false alarm}, or its reciprocal, the \emph{false alarm rate}. Detection delay is generally measured by considering the \emph{expected detection delay},  conditioned on the change-point and possibly the history of observations before the change-point, and taking the supremum over all possible change-points over an infinite horizon. However, these metrics for false alarm and detection delay might not be appropriate in many applications. 
Our work is inspired by the regret analysis for piecewise stationary bandits in \cite{besson2022efficient}, in which 
the metrics used for detecting changes in the bandit environment are the \emph{probability of false alarm} and the \emph{probability of detection delay exceeding a threshold}. Furthermore, the false alarm and delay events are only relevant over a \emph{finite horizon}, and the horizon is not known to the change detector. 



We therefore pose a variant of the QCD optimization problem in which the goal is to minimize the delay threshold (\emph{latency}), under constraints on the probability of false alarm and the probability that the latency is exceeded over a finite horizon, and under the assumption that the horizon is not known to the detector.  
We develop a variant of the cumulative sum (CuSum) test with a time-varying threshold (TVT) as a candidate solution to the optimization problem. We further develop a lower bound on the minimum value of the latency under the constraints. We use the lower bound to establish certain asymptotic optimality properties of the TVT-CuSum test in terms of the horizon and the false alarm probability. We believe that a theoretical study of this variant of the QCD problem will be of relevance to the performance analysis of algorithms for piecewise stationary bandits and reinforcement learning  that employ QCD tests to detect changes in the environment (see, e.g., \cite{liu2018change,cao2019nearly,padakandla2020reinforcement,dahlin2023controlling,wang2021near,zhou2020near,zhou2020nonstationary}).

To the best of our knowledge, there is no prior work on the variant of the QCD problem described in the previous paragraph. A related variant is investigated in \cite{chandar2008optimal}, where it is assumed that the observations are discrete (we do not require this assumption). In \cite{chandar2008optimal}, the desired latency is fixed and assumed to be known to the detector, and the goal is to maximize the horizon. 

The remainder of the paper is organized as follows: The formulation of our QCD problem is given in Section \ref{sec:ProbForm}. 
Section~\ref{sec:CuSum} is devoted to the performance analysis of the TVT-CuSum test. In Section~\ref{sec:LowBound}, we provide a lower bound, which is useful in establishing some asymptotic optimality properties of the TVT-CuSum test. 
Numerical results that validate the analysis are given in Section \ref{sec:sim}, and some concluding remarks are give in Section \ref{sec:Sum}.

\section{High Probability Low Latency Quickest Change Detection}

\label{sec:ProbForm}

Let $\{X_n: n \in \mbb{N}\}$ be a sequence of independent random vectors whose values are observed sequentially.  For the change-point $\nu\in\mbb{N}$, 
%
\begin{align}
    X_{n}\sim\begin{dcases}
    f_{0},\;n < \nu\\
    f_{1},\;n\geq \nu
    \end{dcases}.\label{eq:sample_distr}
\end{align}
In other words, before the change-point $\nu$, the observations follow the pre-change distribution with density $f_{0}$ with respect to some dominating measure $\lambda$. The remaining observations follow the post-change distribution with density $f_{1}$ with respect to the same dominating measure $\lambda$. We use $\Pr_{\nu}$ to denote the probability measure when the change-point occurs at $\nu\in\mbb{N}$, and $\Pr_{\infty}$ to denote the probability measure when there is no change-point (i.e., $\nu= \infty$).

For horizon $T\in\mbb{N}$, we observe the random vectors $X_{1},\dots,X_{T}$ sequentially. Let $\tau$ be the \emph{stopping time} of a (causal) change detector. Our goal is to minimize the latency $d$, for which $\Pr_{\nu}\lp\tau\geq\nu+d\rp$ is small for all $\nu\in\lbp1,\dots,T-d\rbp$, while the probability of false alarm over the horizon $\Pr_{\infty}\lp\tau\leq T\rp$ is small as well. Therefore, the optimization problem of interest has the following form:
\begin{equation}
\begin{split}
    &\underset{\tau}{\textrm{minimize}} \quad d\\
    &\;\textrm{s.t.}\enspace\quad\Pr_{\infty}\lp\tau\leq T\rp\leq\delta_{\mathrm{F}}\\
&\quad\quad\quad\Pr_{\nu}\lp\tau\geq\nu+d\rp\leq\delta_{\mathrm{D}},\;\forall\,\nu\in\lbp1,\dots,T-d\rbp
\end{split}
\label{eq:QCD}
\end{equation}
where $\delta_{\mathrm{F}},\delta_{\mathrm{D}}\in\lp0,1\rp$ are some (small) numbers. In addition, we require change detector (equivalently $\tau$) to be oblivious to the knowledge of the horizon $T$.

\section{Performance of CUSUM Detector}

\label{sec:CuSum}
%

For the standard Lorden formulation of the QCD problem \cite{lorden1971procedures}, the CuSum test is known to be optimal \cite{moustakides1986optimal}. The CuSum test statistic is given by:
\begin{equation}\label{eq:TVT_CuSum_statistic}
    W_n = \max_{1\leq k \leq n} \sum_{i=k}^n \log \frac{f_1(X_i)}{f_0(X_i)} 
\end{equation}
which satisfies the recursion
\begin{equation}    \label{eq:TVT_CuSum_recursion}
 W_n   = \max\{W_{n-1},0\} + \log \frac{f_1(X_n)}{f_0(X_n)}
\end{equation}
with $W_0=0$. The CuSum stopping time is given by:
\begin{equation}
\label{def:cusum_st}
    \tilde{\tau}_b \coloneqq \inf \{n \in \mbb{N}:W_n\geq b \}.
\end{equation}
It is therefore natural to consider the CuSum test of \eqref{def:cusum_st} as a possible candidate for solving the problem of interest given in \eqref{eq:QCD}. However, for the CuSum test with a constant threshold $b$, the false alarm probability $\Pr_{\infty}\lp \tilde{\tau}_b<T\rp$ goes to $1$ as the horizon $T$ goes to infinity. This is because  $\E_{\infty}\lb\tilde{\tau}_b\rb<\infty$ (see, e.g., \cite{lorden1971procedures}). Therefore, the false alarm probability cannot be controlled to a given level $\delta_{\mathrm{F}}$ without the knowledge of $T$.
%
%

Taking a cue from the analysis in \cite{kaufmann2021mixture}, we will show that if we let the threshold $b$ in \eqref{def:cusum_st} increase logarithmically with $n$, the false alarm probability remains upper bounded by a constant for all $T\in\mbb{N}$. This leads to the following modified version of the CuSum test $\tau_{r}$, which we refer to as the time-varying threshold CuSum (TVT-CuSum) test:
%
%
%
\begin{align}
    \tau_r \coloneqq \inf\lbp n\in\mbb{N}:\;W_n \geq\beta\lp n,\delta_{\mathrm{F}},r\rp\rbp,\;r>1\label{eq:TVT_CuSum}
\end{align}
where 
%
%
\begin{equation}
    \beta\lp n,\delta_{\mathrm{F}},r\rp\coloneqq\log\lp\zeta\lp r\rp\frac{n^{r}}{\delta_{\mathrm{F}}}\rp\label{eq:threshold}
\end{equation}
with $\zeta\lp r\rp \coloneqq\sum_{i=1}^{\infty}\frac{1}{i^{r}}$. Then, the following theorem shows that the TVT-CuSum test satisfies the false alarm probability constraint. Note that $\beta$ is not a function of the horizon $T$.

\begin{theorem}[TVT-CuSum Test: False Alarm Probability]\label{prop:FA} For any horizon $T\in\mbb{N}$ and $r>1$, 
%
\begin{equation}
    \Pr_{\infty}\lp\tau_r\leq T\rp\leq\delta_{\mathrm{F}}.
\end{equation}
\end{theorem}

\begin{proof}

First, we can upper bound the probability of false alarm as follows: For all $T\in\mbb{N}$
\begin{align}\nonumber
    &\Pr_{\infty}\lp\tau_r\leq T\rp\\\nonumber
    &\leq\Pr_{\infty}\lp\tau_r<\infty\rp\\\nonumber   &=\Pr_{\infty}\lp\exists\,n\in\mbb{N}:\;W_n\geq\beta\lp n,\delta_{\mathrm{F}},r\rp\rp\\\nonumber
   &=\Pr_{\infty}\lp\exists\,j\leq n:\;\sum_{i=j}^{n}\log\lp\frac{f_{1}\lp X_{i}\rp}{f_{0}\lp X_{i}\rp}\rp\geq\log\lp\zeta\lp r\rp\frac{n^{r}}{\delta_{\mathrm{F}}}\rp\rp\\\nonumber
    &=\Pr_{\infty}\lp\exists\,j\leq n:\;\prod_{i=j}^{n}\frac{f_{1}\lp X_{i}\rp}{f_{0}\lp X_{i}\rp}\geq\zeta\lp r\rp\frac{n^{r}}{\delta_{\mathrm{F}}}\rp\\\nonumber
&=\Pr_{\infty}\Bigg(\exists\,j\in\mbb{N},\;\exists\,k\in\mbb{N}\cup\lbp0\rbp:\\\nonumber
&\quad\quad\quad\quad\;\prod_{i=j}^{j+k}\frac{f_{1}\lp X_{i}\rp}{f_{0}\lp X_{i}\rp}\geq\zeta\lp r\rp\frac{\lp j+k\rp^{r}}{\delta_{\mathrm{F}}}\Bigg)\\\nonumber
&=\Pr_{\infty}\Bigg(\exists\,j\in\mbb{N},\;\exists\,k\in\mbb{N}\cup\lbp0\rbp:\\
&\quad\quad\quad\quad\;\;\frac{1}{\lp j+k\rp^{r}}\prod_{i=j}^{j+k}\frac{f_{1}\lp X_{i}\rp}{f_{0}\lp X_{i}\rp}\geq\frac{\zeta\lp r\rp}{\delta_{\mathrm{F}}}\Bigg)\label{eq:prop_sketch_1}.
\end{align}
Next, from \eqref{eq:prop_sketch_1} we obtain:
\begin{align}\nonumber
    &\Pr_{\infty}\lp\tau_{r}\leq T\rp\\\nonumber
    &\overset{(a)}{\leq}\sum_{j=1}^{\infty}\Pr_{\infty}\lp\exists\,k\in\mbb{N}\cup\lbp0\rbp:\;\frac{1}{\lp j+k\rp^{r}}\prod_{i=j}^{j+k}\frac{f_{1}\lp X_{i}\rp}{f_{0}\lp X_{i}\rp}\geq\frac{\zeta\lp r\rp}{\delta_{\mathrm{F}}}\rp\\\nonumber
    &\overset{(b)}{\leq}\sum_{j=1}^{\infty}\frac{\delta_{\mathrm{F}}}{\zeta\lp r\rp}\E_{\infty}\lb\frac{1}{j^{r}}\frac{f_{1}\lp X_{j}\rp}{f_{0}\lp X_{j}\rp}\rb =\frac{\delta_{\mathrm{F}}}{\zeta\lp r\rp}\sum_{j=1}^{\infty}\frac{1}{j^{r}} =\delta_{\mathrm{F}}
\end{align}
where step $(a)$ results from union bound. Step $(b)$ stems from Doob's submartingale inequality \cite{doob1953stochastic} since $\lp\frac{1}{\lp j+k\rp^{r}}\prod_{i=j}^{j+k}\frac{f_{1}\lp X_{i}\rp}{f_{0}\lp X_{i}\rp}\rp_{k=0}^{\infty}$ is a supermartingale; this is shown in Lemma \ref{lem:sup_martin}, which, along with its proof, is given in Appendix~\ref{sec:lem1}.
\end{proof}

Next, we analyze the high probability latency for the TVT-CuSum test, which is defined as follows:
\begin{align}\nonumber
    &d_r\lp T,\delta_{\mathrm{F}},\delta_{\mathrm{D}}\rp\\
    &\coloneqq \inf\lbp d\in\mbb{N}:\;\Pr_{\nu}\lp\tau_r\geq\nu+d\rp\leq\delta_{\mathrm{D}}\;\forall\,\nu\in\lbp1,\dots,T-d\rbp\rbp.\label{eq:HPDD-CuSum}
\end{align}
For the purpose of our analysis, we define $\Lambda$ to be the cumulant generating function of $\log\lp\frac{f_{0}\lp X\rp}{f_{1}\lp X\rp}\rp$ with $X\sim f_{1}$, i.e., 
\begin{equation} \Lambda\lp\theta\rp=\log\lp\E_{f_{1}}\lb\exp\lp\theta\log\lp\frac{f_{0}\lp X\rp}{f_{1}\lp X\rp}\rp\rp\rb\rp.\label{eq:cumulant}
\end{equation}
%
The following theorem gives an upper bound on $d_{r}$.

\begin{theorem}[High Probability Latency for TVT-CuSum Test]\label{prop:HPDD_CuSum} For all $T\in\mbb{N}$, $\delta_{\mathrm{F}},\delta_{\mathrm{D}}\in\lp0,1\rp$, $r>1$, 
\begin{align}\nonumber
    &d_r\lp T,\delta_{\mathrm{F}},\delta_{\mathrm{D}}\rp\\\nonumber
    &\leq\inf_{\theta\in\lp0,1\rp}\bigg\{\frac{1}{\lvert\Lambda\lp\theta\rp\rvert}\bigg[\log\lp\frac{1}{\delta_{\mathrm{D}}}\rp+\theta\log\lp\frac{1}{\delta_{\mathrm{F}}}\rp\\
    &\quad\quad\quad\quad\quad\quad\quad\quad\;+r\theta\log\lp T\rp+\theta\log\lp\zeta\lp r\rp\rp\bigg]\bigg\}.\label{eq:HPDD_upp}
\end{align}
    
\end{theorem}

\begin{proof}

Fix an arbitrary $T\in\mbb{N}$ and arbitrary $\delta_{\mathrm{F}},\delta_{\mathrm{D}}\in\lp0,1\rp$. 
First, it is easy to show that $\Lambda\lp 0 \rp=\Lambda\lp1\rp=0$.
%
%
Since $f_{1}\neq f_{0}$,  $\Lambda$ is strictly convex, and therefore, $\Lambda\lp\theta\rp<0$ $\forall\,\theta\in\lp0,1\rp$. Next, by the definition of $\tau_{r}$, $\forall\,d\in\mbb{N}$ and $\forall\,\nu\in\lbp1,\dots,T-d\rbp$
\begin{align}\nonumber
    &\Pr_{\nu}\lp\tau_r\geq\nu+d\rp\\\nonumber
    &=\Pr_{\nu}\lp\inf\lbp n\in\mbb{N}:\;W_n\geq\beta\lp n,\delta_{\mathrm{F}},r\rp\rbp\geq\nu+d\rp\\\nonumber
    &=\Pr_{\nu}\Bigg(\forall\,n\in\lbp1,\dots,\nu+d-1\rbp:\\
    &\quad\quad\quad\;\max_{1\leq j\leq n}\sum_{i=j}^{n}\log\lp\frac{f_{1}\lp X_{i}\rp}{f_{0}\lp X_{i}\rp}\rp<\log\lp\zeta\lp r\rp\frac{n^{r}}{\delta_{\mathrm{F}}}\rp\bigg).\label{eq:prop_2_sketch_1}
\end{align}
Then, we have $\forall\,\theta\in\lp0,1\rp$
\begin{align}
&\Pr_{\nu}\lp\tau_r\geq\nu+d\rp\\\nonumber
&\overset{(a)}{\leq}\Pr_{\nu}\Bigg(\max_{1\leq j\leq\nu+d-1}\sum_{i=j}^{\nu+d-1}\log\lp\frac{f_{1}\lp X_{i}\rp}{f_{0}\lp X_{i}\rp}\rp\nonumber\\
&\quad\quad\quad\quad<\log\lp\zeta\lp r\rp\frac{\lp\nu+d\rp^{r}}{\delta_{\mathrm{F}}}\rp\Bigg)\nonumber\\
&\overset{(b)}{\leq}\Pr_{\nu}\lp\sum_{i=\nu}^{\nu+d-1}\log\lp\frac{f_{1}\lp X_{i}\rp}{f_{0}\lp X_{i}\rp}\rp<\log\lp\zeta\lp r\rp\frac{\lp\nu+d\rp^{r}}{\delta_{\mathrm{F}}}\rp\rp\nonumber\\
&=\Pr_{\nu}\lp-\sum_{i=\nu}^{\nu+d-1}\log\lp\frac{f_{1}\lp X_{i}\rp}{f_{0}\lp X_{i}\rp}\rp>-\log\lp\zeta\lp r\rp\frac{\lp\nu+d\rp^{r}}{\delta_{\mathrm{F}}}\rp\rp\nonumber\\
&\overset{(c)}{\leq}\exp\lp-d\lp-\frac{1}{d}\theta\log\lp\zeta\lp r\rp\frac{\lp\nu+d\rp^{r}}{\delta_{\mathrm{F}}}\rp-\Lambda\lp\theta\rp\rp\rp\nonumber\\
&=\lp\zeta\lp r\rp\frac{\lp\nu+d\rp^{r}}{\delta_{\mathrm{F}}}\rp^{\theta}\exp\lp d\Lambda\lp\theta\rp\rp \label{eq:ub_nu}\\
&\overset{(d)}{\leq}\lp\zeta\lp r\rp\frac{T^{r}}{\delta_{\mathrm{F}}}\rp^{\theta}\exp\lp d\Lambda\lp\theta\rp\rp\label{eq:prop_2_sketch_2}
\end{align}
where step $(a)$ is due to the fact that $\lbp \nu+d\rbp\subseteq\lbp1,\dots,\nu+d\rbp$, while step $(b)$ is owing to the fact that $\sum_{i=\nu}^{\nu+d-1}\log\lp\frac{f_{1}\lp X_{i}\rp}{f_{0}\lp X_{i}\rp}\rp\leq\max_{1\leq j\leq\nu+d-1}\sum_{i=j}^{\nu+d-1}\log\lp\frac{f_{1}\lp X_{i}\rp}{f_{0}\lp X_{i}\rp}\rp$. Step $(c)$ stems from the Chernoff bound \cite{chernoff1952measure}, and step $(d)$ is due to the fact that $\nu+d\leq T$.

Define
\begin{align}\nonumber
    \tilde{d}\coloneqq\frac{1}{\lvert\Lambda\lp\theta\rp\rvert}\bigg[&\log\lp\frac{1}{\delta_{\mathrm{D}}}\rp+\theta\log\lp\frac{1}{\delta_{\mathrm{F}}}\rp\\
    &+r\theta\log\lp T\rp+\theta\log\lp\zeta\lp r\rp\rp\bigg].
\end{align}
Then, according to \eqref{eq:prop_2_sketch_2}, we have:
\begin{align}\nonumber
    \Pr_{\nu}\lp\tau_r>\nu+\tilde{d}\rp&\leq\lp\zeta\lp r\rp\frac{T^{r}}{\delta_{\mathrm{F}}}\rp^{\theta}\exp\lp \tilde{d}\Lambda\lp\theta\rp\rp\\
    &=\delta_{\mathrm{D}}.\label{eq:prop_2_sketch_3}
\end{align}
By the definition of $d_r$, $d_r\leq\tilde{d}$, and thus  \eqref{eq:HPDD_upp} holds.
\end{proof}


Theorem \ref{prop:HPDD_CuSum} shows that $d_r=\mcal{O}\lp\log T\rp$. In the next section, we demonstrate that this growth of $d_r$ with $T$ is order optimal when $T$ is large and $\delta_{\mathrm{F}}+\delta_{\mathrm{D}}<1$. 

\section{Asymptotic Information-Theoretic Lower Bound on the High Probability Detection Delay}

\label{sec:LowBound}

Let $\tau^{*}$ be the solution to \eqref{eq:QCD} and $d^{*}\lp T,\delta_{\mathrm{F}},\delta_{\mathrm{D}}\rp$ be the corresponding minimum value. In this section, we present a lower bound for $d^{*}\lp T,\delta_{\mathrm{F}},\delta_{\mathrm{D}}\rp$. For the purpose of our analysis, we define the following constant
\begin{align}
C\coloneqq\log\lp\E_{f_{1}}\lb\frac{f_{1}\lp X\rp}{f_{0}\lp X\rp}\rb\rp.\label{eq:C_const}   
\end{align}
It is easy to see that $C>0$ using Jensen's inequality. Furthermore, we assume that $C<\infty$.

\begin{theorem}[Lower Bound for High Probability Latency]\label{thm:low_bound_HPDD}For all $\delta_{\mathrm{F}},\delta_{\mathrm{D}}\in\lp0,1\rp$ such that $\delta_{\mathrm{F}}+\delta_{\mathrm{D}}<1$
\begin{align}\nonumber
&d^{*}\lp T,\delta_{\mathrm{F}},\delta_{\mathrm{D}}\rp\geq \lp\frac{1}{C}+o\lp 1\rp\rp \\\nonumber
&
\quad \cdot \lb\log\lp T\rp+\log\lp\frac{1}{\delta_{\mathrm{F}}}\rp+\log\lp1-\delta_{\mathrm{F}}-\delta_{\mathrm{D}}\rp+o\lp1\rp\rb\label{eq:d_lower}
\end{align}
as $T\to \infty$.
\end{theorem}

\begin{proof} 
%
%

For $c>C$, define the events
\begin{equation}
\mcal{A}\coloneqq\lbp\nu\leq\tau^{*}<\nu+d^{*},\sum_{i=\nu}^{\nu+d^{*}-1}\!\!\!\!\log\lp\frac{f_{1}\lp X_{i}\rp}{f_{0}\lp X_{i}\rp}\rp\geq d^{*}c\rbp
\end{equation}
and 
\begin{equation}\label{eq:Bc}
\mcal{B}\coloneqq\lbp\nu\leq\tau^{*}<\nu+d^{*},\sum_{i=\nu}^{\nu+d^{*}-1}\!\!\!\!\log\lp\frac{f_{1}\lp X_{i}\rp}{f_{0}\lp X_{i}\rp}\rp<d^{*}c\rbp.
\end{equation}
We note that $\mcal{A}\cap \mcal{B} = \emptyset$ and $\mcal{A}\cup \mcal{B}= \lbp \nu\leq \tau^{*} < \nu+d^{*}\rbp$, which we will use later in the proof.

From the problem formulation \eqref{eq:QCD} we have that 
%
$\forall\,\nu\in\lbp1,\dots,T-d^{*}\rbp$, 
\begin{align}\nonumber
\delta_{\mathrm{D}}&\geq\Pr_{\nu}\lp\tau^{*}\geq\nu+d^{*}\rp\\\nonumber
&=1-\Pr_{\nu}\lp\tau^{*}<\nu+d^{*}\rp\\\nonumber
&=1-\Pr_{\nu}\lp\tau^{*}<\nu\rp-\Pr_{\nu}\lp\nu\leq\tau^{*}<\nu+d^{*}\rp\\\nonumber
&=1-\Pr_{\infty}\lp\tau^{*}<\nu\rp-\Pr_{\nu}\lp\nu\leq\tau^{*}<\nu+d^{*}\rp\\
&\geq1-\Pr_{\infty}\lp\tau^{*}\leq T\rp-\Pr_{\nu}\lp\mcal{B}\rp-\Pr_{\nu}\lp\mcal{A}\rp.\label{eq:thm_1_sketch_1}
\end{align}
Next, since $\tau^{*}$ satisfies the false alarm probability constraint $\Pr_{\infty}\lp\tau^{*}\leq T\rp\leq\delta_{\mathrm{F}}$, by Lemma \ref{lem:LowBoundFA} in Appendix \ref{sec:lem2}, there exists a change-point $\tilnu \in\lbp1,\dots,T-d^{*}\rbp$ such that
\begin{align}
    \Pr_{\infty}\lp\tilnu\leq\tau^* < \tilnu+d^{*}\rp\leq\frac{\delta_{\mathrm{F}}}{\lfl T/d^*\rfl}.\label{eq:FA_window_tilnu}
\end{align}
For this choice of $\tilnu$, we have
\begin{align}\nonumber
\Pr_{\tilnu}\lp\mcal{B}\rp&=\int_{\mcal{B}}\prod_{i=1}^{\tilnu-1}f_{0}\lp x_{i}\rp\prod_{i=\tilnu}^{\tilnu+d^{*}-1}f_{1}\lp x_{i}\rp\otimes_{i=1}^{\tilnu+d^{*}-1} d \lambda\lp x_{i}\rp\\\nonumber  &=\int_{\mcal{B}}\prod_{i=\tilnu}^{\tilnu+d^{*}-1}\frac{f_{1}\lp x_{i}\rp}{f_{0}\lp x_{i}\rp}\prod_{i=1}^{\tilnu+d^{*}-1}f_{0}\lp x_{i}\rp\otimes_{i=1}^{\tilnu+d^{*}-1} d \lambda\lp x_{i}\rp\\\nonumber
&\overset{(a)}{\leq}\int_{\mcal{B}}\exp\lp d^{*}c\rp\prod_{i=1}^{\tilnu+d^{*}-1}f_{0}\lp x_{i}\rp\otimes_{i=1}^{\tilnu+d^{*}-1} d \lambda\lp x_{i}\rp\\\nonumber
&=\exp\lp d^{*}c\rp\int_{\mcal{B}}\prod_{i=1}^{\tilnu+d^{*}-1}f_{0}\lp x_{i}\rp\otimes_{i=1}^{\tilnu+d^{*}-1} d \lambda\lp x_{i}\rp\\\nonumber
&\overset{(b)}{=}\exp\lp d^{*}c\rp\Pr_{\infty}\lp\mcal{B}\rp\\\nonumber
&\overset{(c)}{\leq}\exp\lp d^{*}c\rp\Pr_{\infty}\lp\tilnu\leq\tau^{*}<\tilnu+d^{*}\rp\\
&\overset{(d)}{\leq}\frac{\exp\lp d^{*}c\rp\delta_{\mathrm{F}}}{\lfl T/d^{*}\rfl}\label{eq:thm_1_sketch_2}
\end{align}
where step $(a)$ results from the definition of $\mcal{B}$, and step $(b)$ stems from the fact that under $\Pr_{\infty}$, every $X_{i}$ follows the density $f_{0}$. Step $(c)$ is owing to the fact that $\mcal{B}\subseteq\lbp\tilnu\leq\tau^{*}<\tilnu+d^{*}\rbp$, and step $(d)$ is due to \eqref{eq:FA_window_tilnu}.

Then, for any $\nu\in\lbp1,\dots,T-d^{*}\rbp$,
\begin{align}\nonumber
&\Pr_{\nu}\lp\mcal{A}\rp\\\nonumber
&=\Pr_{\nu}\lp\nu\leq\tau^{*}<\nu+d^{*},\sum_{i=\nu}^{\nu+d^{*}-1}\log\lp\frac{f_{1}\lp X_{i}\rp}{f_{0}\lp X_{i}\rp}\rp\geq d^{*}c\rp\\\nonumber
&\leq\Pr_{\nu}\lp\sum_{i=\nu}^{\nu+d^{*}-1}\log\lp\frac{f_{1}\lp X_{i}\rp}{f_{0}\lp X_{i}\rp}\rp\geq d^{*}c\rp\\\nonumber
&=\Pr_{\nu}\lp\prod_{i=\nu}^{\nu+d^{*}-1}\frac{f_{1}\lp X_{i}\rp}{f_{0}\lp X_{i}\rp}\geq e^{d^{*}c}\rp\\\nonumber
&\overset{(a)}{\leq}e^{-d^{*}c}\E_{\nu}\lb\prod_{i=\nu}^{\nu+d^{*}-1}\frac{f_{1}\lp X_{i}\rp}{f_{0}\lp X_{i}\rp}\rb\\\nonumber
&=e^{-d^{*}c}\prod_{i=\nu}^{\nu+d^{*}-1}\E_{\nu}\lb\frac{f_{1}\lp X_{i}\rp}{f_{0}\lp X_{i}\rp}\rb\\
&\overset{(b)}{\leq}e^{-d^{*}\lp c-C\rp}\label{eq:thm_1_sketch_3}
\end{align}
where step $(a)$ results from the Markov inequality.
Step $(b)$ is due to the definition of $C$, which is in the theorem statement.

Next, by plugging \eqref{eq:thm_1_sketch_2} and \eqref{eq:thm_1_sketch_3} into \eqref{eq:thm_1_sketch_1} with $\nu=\tilnu$, we have 
\begin{align}\nonumber
\delta_{\mathrm{D}}&\geq1-\delta_{\mathrm{F}}-\frac{\exp\lp d^{*}c\rp\delta_{\mathrm{F}}}{\lfl T/d^{*}\rfl}-e^{-d^{*}\lp c-C\rp}\\
&\geq 1-\delta_{\mathrm{F}}-\frac{d^{*}\exp\lp d^{*}c\rp\delta_{\mathrm{F}}}{T-d^{*}}-e^{-d^{*}\lp c-C\rp}.\label{eq:thm_1_sketch_4}
\end{align}

By Theorem \ref{prop:FA} and \ref{prop:HPDD_CuSum}, $d^{*}\leq d_{r}$ for any $r>1$. Since $d_{r}$ grows logarithmically with $T$, $d^{*}=\mcal{O}\lp\log\lp T\rp\rp$; therefore, we can observe that for any $b\in\lp0,1\rp$, $d^{*}\leq bT$ for $T$ large enough. In addition, this $b$ approaches $0$ as $T$ goes to infinity. Thus, for any $s\in\lp0,1\rp$, $T-d^{*}\geq sT$ for $T$ large enough, and $s$ approaches $1$ as $T$ goes to infinity. 

Furthermore, $\delta_{\mathrm{F}}+\delta_{\mathrm{D}}<1$, $d^{*}\lp T,\delta_{\mathrm{F}},\delta_{\mathrm{D}}\rp\rightarrow\infty$ as $T\rightarrow\infty$ by Lemma \ref{lem:d_inf} in Appendix \ref{sec:lem4}. Now, we choose $c\rightarrow C$ such that $e^{-d^{*}\lp c-C\rp}\rightarrow0$ as $T\rightarrow\infty$. 

Again, since $d^{*}\rightarrow\infty$ as $T\rightarrow\infty$, for any $\tilde{c}>c$, $d^{*}\exp\lp cd^{*}\rp<\exp\lp \tilde{c}d^{*}\rp$ for $T$ large enough, and $\tilde{c}$ goes to $c$ as $T$ goes to infinity. Hence, by \eqref{eq:thm_1_sketch_4}, for $T$ large enough, we have:
\begin{align}
\delta_{\mathrm{D}}\geq 1-\delta_{\mathrm{F}}-\frac{\delta_{\mathrm{F}}}{sT}\exp\lp\tilde{c}d^{*}\rp-\epsilon\label{eq:thm_1_sketch_5}
\end{align}
where $\epsilon\in\lp0,1-\delta_{\textrm{F}}-\delta_{\textrm{D}}\rp$, $\epsilon\rightarrow0$ as $T\rightarrow\infty$. Rearranging the terms in \eqref{eq:thm_1_sketch_5}, we obtain
\begin{align}
\frac{\delta_{\mathrm{F}}}{sT}\exp\lp\tilde{c}d^{*}\rp\geq1-\delta_{\mathrm{F}}-\delta_{\mathrm{D}}-\epsilon.\label{eq:thm_1_sketch_6}
\end{align}
By taking $\log$ on both sides, we have
\begin{align}\nonumber
\tilde{c}d^{*}\geq&\log\lp T\rp\\
&+\log\lp\frac{1}{\delta_{\mathrm{F}}}\rp+\log\lp s\rp+\log\lp1-\delta_{\mathrm{F}}-\delta_{\mathrm{D}}-\epsilon\rp. \label{eq:thm_1_sketch_7}
\end{align}
This implies that
\begin{align}\nonumber
d^{*}\lp T,\delta_{\mathrm{F}},\delta_{\mathrm{D}}\rp\geq&\frac{1}{\tilde{c}}\log\lp T\rp+\frac{1}{\tilde{c}}\log\lp\frac{1}{\delta_{\mathrm{F}}}\rp\\
&+\frac{1}{\tilde{c}}\log\lp s\rp+\frac{1}{\tilde{c}}\log\lp1-\delta_{\mathrm{F}}-\delta_{\mathrm{D}}-\epsilon\rp.
\end{align}
Since $\tilde{c}\rightarrow c$, $c\rightarrow C$, $s\rightarrow1$, and $\epsilon\rightarrow0$, all as $T\rightarrow\infty$, we have the theorem.
\end{proof}

From Theorem \ref{thm:low_bound_HPDD}, it is straightforward to see the following corollary.

\begin{corollary}
    When $\delta_{\mathrm{D}}+\delta_{\mathrm{F}}<1$, as $T$ goes to infinity, $d^{*}\lp T,\delta_{\mathrm{F}},\delta_{\mathrm{D}}\rp\geq\frac{1}{C}\log\lp T\rp\lp1+o(1)\rp$
\end{corollary}

This corollary, along with Theorem~\ref{prop:HPDD_CuSum}, shows that the growth of $d_{r}$ with $T$ of the TVT-CuSum test in \eqref{eq:TVT_CuSum} is order optimal when $T$ is large and $\delta_{\mathrm{D}}+\delta_{\mathrm{F}}<1$.

\section{Simulation}

\label{sec:sim}

\begin{figure}
    \centering
    \includegraphics[width=9cm]{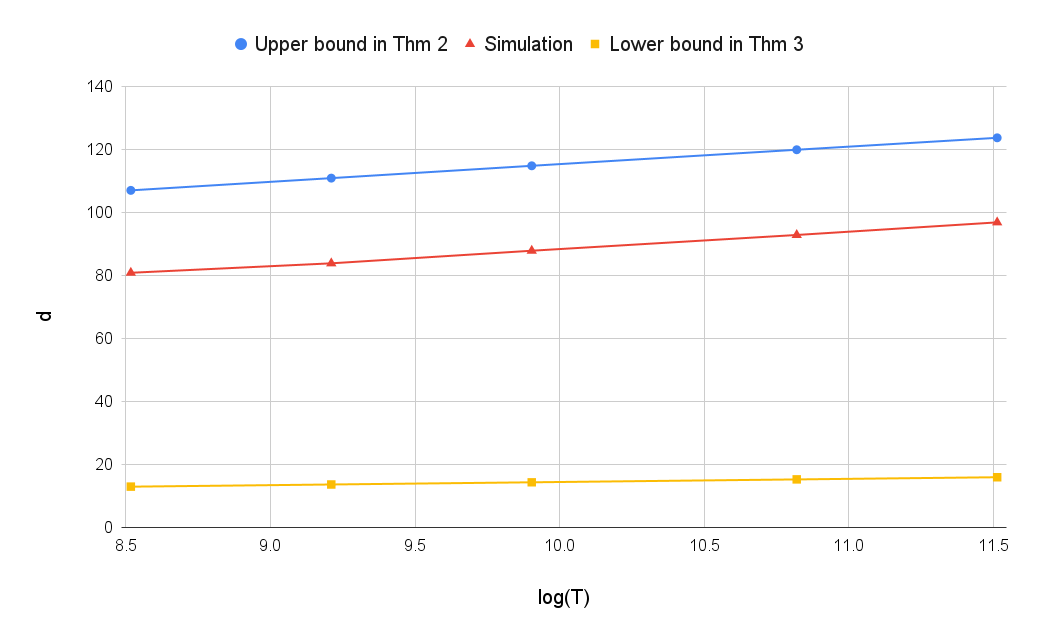}
    \caption{High probability latency $d_r$ as a function of the horizon $T$, with $\delta_{\mathrm{F}}=\delta_{\mathrm{D}} =0.01$.}
    \label{fig:HPDD-T}
\end{figure}

\begin{figure}
    \centering
    \includegraphics[width=9cm]{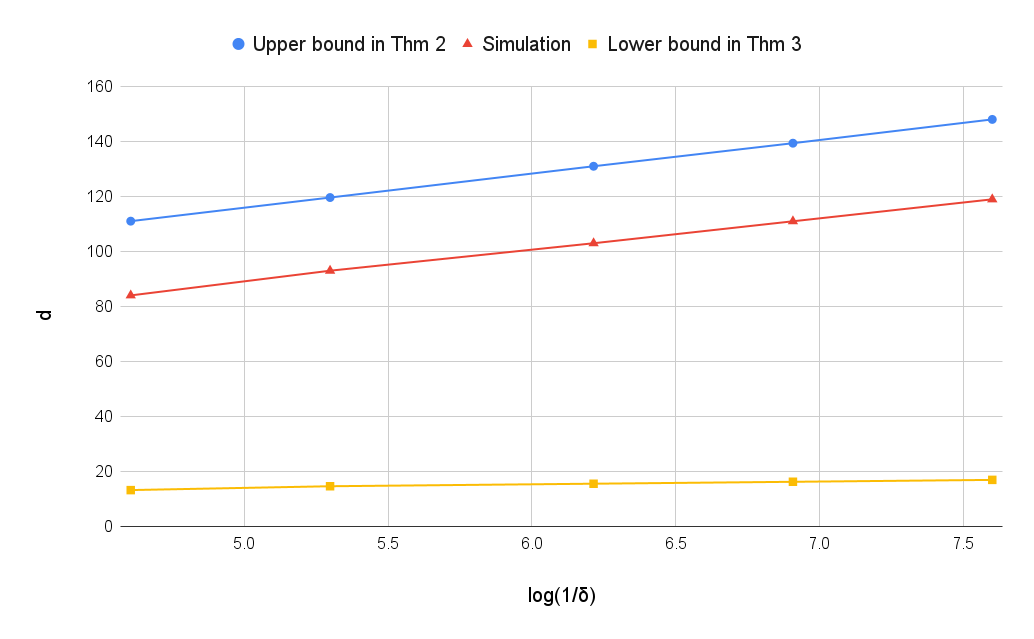}
    \caption{High probability latency $d_{r}$ as a function of $\delta = \delta_{\mathrm{F}} = \delta_{\mathrm{D}}$, with the horizon $T=10000$}
    \label{fig:HPDD-delta}
\end{figure}

In this section, we present some simulation results for the performance of the TVT-CuSum test and compare it with the upper bound in Theorem \ref{prop:HPDD_CuSum} and lower bound in Theorem \ref{thm:low_bound_HPDD}. The number of trials used was $200000$. The pre-change distribution was taken to be $\mcal{N}\lp0,1\rp$, while the post-change distribution was taken to be $\mcal{N}\lp1,1\rp$. According to \eqref{eq:ub_nu}, an upper bound on the probability of the delay exceeding the high probability latency $d$ increases\footnote{This is in contrast with the standard CuSum test with a constant threshold, for which the worst-case delay occurs when the change-point is at $\nu=0$.} with $\nu$. 
Hence, to simulate the worst-case scenario for the high probability latency across $\nu$, we should pick a value of $\nu$ closer to the horizon $T$ than 0. We therefore set $\nu$ to be difference between the horizon $T$ and the upper bound in Theorem \ref{prop:HPDD_CuSum}. The parameter $r$ of the TVT-CuSum test is set to be $2$ in the experiments. If the test failed to detect the change before the horizon, which was very rare in our simulations,  we set the delay to be $T-\nu$. We recorded the detection delay for each trial and picked the $100(1-\delta_{\mathrm{D}})^{\mathrm{th}}$ percentile out of all the values to be the empirical value of the high probability latency $d$. Then, we compared it with the upper bound in Theorem \ref{prop:HPDD_CuSum} and lower bound in Theorem \ref{thm:low_bound_HPDD}. In the experiments, the quantities $\delta_{\mathrm{F}}$ and $\delta_{\mathrm{D}}$ are set to the same value, denoted by $\delta$. Figure \ref{fig:HPDD-T} and \ref{fig:HPDD-delta} exhibit how the high probability latency grows with $T$ and $\delta$

As shown in Figure \ref{fig:HPDD-T} and \ref{fig:HPDD-delta}, the upper bound is closer to the simulation results than the lower bound, and the slope of the upper bound and that of the simulation results are approximately the same. 

The figures also indicate that the lower bound is very loose. There may two reasons why this the case: (i) in deriving the lower bound, we did not impose the condition that test $\tau$ is oblivious to the knowledge of the horizon $T$; and (ii) the TVT-CuSum test could be improved considerably. We believe that the first reason is more likely to be true.


\section{Conclusion}

\label{sec:Sum}

We posed a new variant of the QCD problem with a finite horizon, in which the goal is to minimize the latency, under constraints on the probability of false alarm and the probability of the latency being exceeded. We proposed the TVT-CuSum test with a threshold that increases logarithmically with time as a candidate solution. We show that the growth of the high probability latency $d_r$ of the TVT-CuSum test with $T$ is order optimal, when $T$ is large and $\delta_{\mathrm{F}}+\delta_{\mathrm{D}}<1$. As indicated in Section \ref{sec:sim}, there might still be room for improvement for the lower bound given in Theorem \ref{thm:low_bound_HPDD}. We leave this refinement of the lower bound for future work. It is also of interest to generalize the analysis in this paper to the more practical scenario in which there is uncertainty in the pre- and post- change distributions.

\printbibliography

@article{besson2022efficient,
  title={Efficient change-point detection for tackling piecewise-stationary bandits},
  author={Besson, Lilian and Kaufmann, Emilie and Maillard, Odalric-Ambrym and Seznec, Julien},
  journal={The Journal of Machine Learning Research},
  volume={23},
  number={1},
  pages={3337--3376},
  year={2022},
  publisher={JMLRORG}
}

@inproceedings{wang2021near,
  title={Near-optimal algorithms for piecewise-stationary cascading bandits},
  author={Wang, Lingda and Zhou, Huozhi and Li, Bingcong and Varshney, Lav R and Zhao, Zhizhen},
  booktitle={ICASSP 2021-2021 IEEE International Conference on Acoustics, Speech and Signal Processing (ICASSP)},
  pages={3365--3369},
  year={2021},
  organization={IEEE}
}

@inproceedings{zhou2020near,
  title={A near-optimal change-detection based algorithm for piecewise-stationary combinatorial semi-bandits},
  author={Zhou, Huozhi and Wang, Lingda and Varshney, Lav and Lim, Ee-Peng},
  booktitle={Proceedings of the AAAI Conference on Artificial Intelligence},
  volume={34},
  number={04},
  pages={6933--6940},
  year={2020}
}

@article{zhou2020nonstationary,
  title={Nonstationary reinforcement learning with linear function approximation},
  author={Zhou, Huozhi and Chen, Jinglin and Varshney, Lav R and Jagmohan, Ashish},
  journal={arXiv preprint arXiv:2010.04244},
  year={2020}
}

@article{xie_vvv_qcd_overview,
	author = {Xie, Liyan and Zou, Shaofeng and Xie, Yao and Veeravalli, Venugopal V.},
	doi = {10.1109/JSAIT.2021.3072962},
	journal = {IEEE Journal on Selected Areas in Information Theory},
	number = {2},
	pages = {494-514},
	title = {Sequential (Quickest) Change Detection: Classical Results and New Directions},
	volume = {2},
	year = {2021}}

@incollection{vvv_qcd_overview,
	address = {Cambridge, MA},
	author = {V. V. Veeravalli and T. Banerjee},
	booktitle = {Academic press library in signal processing: Array and statistical signal processing},
	publisher = {Academic Press},
	title = {Quickest Change Detection},
	year = {2013}}

@book{poor-hadj-qcd-book-2009,
	author = {Poor, H. V. and Hadjiliadis, O.},
	publisher = {Cambridge University Press},
	title = {Quickest detection},
	year = {2009}}

@article{lorden1971procedures,
  title={Procedures for reacting to a change in distribution},
  author={Lorden, Gary},
  journal={The annals of mathematical statistics},
  pages={1897--1908},
  year={1971},
  publisher={JSTOR}
}

@article{kaufmann2021mixture,
  title={Mixture martingales revisited with applications to sequential tests and confidence intervals},
  author={Kaufmann, Emilie and Koolen, Wouter M},
  journal={The Journal of Machine Learning Research},
  volume={22},
  number={1},
  pages={11140--11183},
  year={2021},
  publisher={JMLRORG}
}

@book{tart-niki-bass-2014,
	author = {A. G. Tartakovsky and I. V. Nikiforov and M. Basseville},
	publisher = {CRC Press},
	series = {Statistics},
	title = {Sequential Analysis: {Hypothesis} Testing and Change-Point Detection},
	year = {2014}}

@article{moustakides1986optimal,
	author = {Moustakides, George V.},
	doi = {10.1214/aos/1176350164},
	journal = {Annals of Statistics},
	month = dec,
	number = {4},
	pages = {1379-1387},
	publisher = {The Institute of Mathematical Statistics},
	title = {Optimal Stopping Times for Detecting Changes in Distributions},
	volume = {14},
	year = {1986}}

@book {doob1953stochastic,
    AUTHOR = {Doob, J. L.},
     TITLE = {Stochastic Processes},
 PUBLISHER = {John Wiley \& Sons, Inc., New York; Chapman \& Hall, Ltd., London},
      YEAR = {1953},
     PAGES = {viii+654},
   MRCLASS = {60.0X},
  MRNUMBER = {58896},
MRREVIEWER = {D.\ G.\ Kendall},
}

@article{chernoff1952measure,
  title={A measure of asymptotic efficiency for tests of a hypothesis based on the sum of observations},
  author={Chernoff, Herman},
  journal={The Annals of Mathematical Statistics},
  pages={493--507},
  year={1952},
  publisher={JSTOR}
}

@inproceedings{dahlin2023controlling,
  title={Controlling a markov decision process with an abrupt change in the transition kernel},
  author={Dahlin, Nathan and Bose, Subhonmesh and Veeravalli, Venugopal V},
  booktitle={2023 American Control Conference (ACC)},
  pages={3401--3408},
  year={2023},
  organization={IEEE}
}

@article{padakandla2020reinforcement,
  title={Reinforcement learning algorithm for non-stationary environments},
  author={Padakandla, Sindhu and KJ, Prabuchandran and Bhatnagar, Shalabh},
  journal={Applied Intelligence},
  volume={50},
  pages={3590--3606},
  year={2020},
  publisher={Springer}
}

@inproceedings{liu2018change,
  title={A change-detection based framework for piecewise-stationary multi-armed bandit problem},
  author={Liu, Fang and Lee, Joohyun and Shroff, Ness},
  booktitle={Proceedings of the AAAI Conference on Artificial Intelligence},
  volume={32},
  number={1},
  year={2018}
}

@article{cao2019nearly,
  title={Nearly optimal adaptive procedure for piecewise-stationary bandit: a change-point detection approach},
  author={Cao, Yang and Zheng, Wen and Kveton, Branislav and Xie, Yao},
  journal={AISTATS, Okinawa, Japan},
  year={2019}
}

@article{chandar2008optimal,
  title={Optimal sequential frame synchronization},
  author={Chandar, Venkat and Tchamkerten, Aslan and Wornell, Gregory},
  journal={IEEE Transactions on Information Theory},
  volume={54},
  number={8},
  pages={3725--3728},
  year={2008},
  publisher={IEEE}
}

\onecolumn

\appendices

\section{Proof of Lemma \ref{lem:sup_martin}}

\label{sec:lem1}

\begin{lemma}\label{lem:sup_martin}
$\forall\,r>1$, under the probability measure $\Pr_{\infty}$, i.e., $X_{m}\diid f_{0}$, then $\forall\,n\in\mbb{N}$, the sequence 
\begin{align}
\lp\frac{1}{\lp n+m\rp^{r}}\prod_{i=n}^{n+m}\frac{f_{1}\lp X_{i}\rp}{f_{0}\lp X_{i}\rp}\rp_{m=0}^{\infty}  
\end{align}
is a nonnegative supermartingale.
\end{lemma}

\begin{proof}
We can observe that $\forall\,n\in\mbb{N}$, $\forall\,m\in\mbb{N}\cup\lbp0\rbp$, $\forall\,r>1$:
\begin{align}\nonumber
&\E_{\infty}\lb\frac{1}{\lp n+m\rp^{r}}\prod_{i=n}^{n+m}\frac{f_{1}\lp X_{i}\rp}{f_{0}\lp X_{i}\rp}\bigg|X_{n},\dots,X_{n+m-1}\rb\\\nonumber
&=\frac{1}{\lp n+m\rp^{r}}\\\nonumber
&\quad\enspace\cdot\prod_{i=n}^{n+m-1}\frac{f_{1}\lp X_{i}\rp}{f_{0}\lp X_{i}\rp}\E_{\infty}\lb\frac{f_{1}\lp X_{n+m}\rp}{f_{0}\lp X_{n+m}\rp}\bigg|X_{n},\dots,X_{n+m-1}\rb\\\nonumber
&=\frac{1}{\lp n+m\rp^{r}}\prod_{i=n}^{n+m-1}\frac{f_{1}\lp X_{i}\rp}{f_{0}\lp X_{i}\rp}\E_{\infty}\lb\frac{f_{1}\lp X_{n+m}\rp}{f_{0}\lp X_{n+m}\rp}\rb\\\nonumber
&=\frac{1}{\lp n+m\rp^{r}}\prod_{i=n}^{n+m-1}\frac{f_{1}\lp X_{i}\rp}{f_{0}\lp X_{i}\rp}\int_{\mbb{R}}\frac{f_{1}\lp x\rp}{f_{0}\lp x\rp}f_{0}\lp x\rp d\lambda\lp x\rp\\\nonumber
&=\frac{1}{\lp n+m\rp^{r}}\prod_{i=n}^{n+m-1}\frac{f_{1}\lp X_{i}\rp}{f_{0}\lp X_{i}\rp}\int_{\mbb{R}}f_{1}\lp x\rp d\lambda\lp x\rp\\\nonumber
&=\frac{1}{\lp n+m\rp^{r}}\prod_{i=n}^{n+m-1}\frac{f_{1}\lp X_{i}\rp}{f_{0}\lp X_{i}\rp}\\
&\leq\frac{1}{\lp n+m-1\rp^{r}}\prod_{i=n}^{n+m-1}\frac{f_{1}\lp X_{i}\rp}{f_{0}\lp X_{i}\rp}.\label{eq:sup_mart}
\end{align}
Since each $\frac{1}{\lp n+m\rp^{r}}\prod_{i=n}^{n+m}\frac{f_{1}\lp X_{i}\rp}{f_{0}\lp X_{i}\rp}$ is nonnegative, Lemma \ref{lem:sup_martin} follows from \eqref{eq:sup_mart}.
\end{proof}

\section{Proof of Lemma \ref{lem:LowBoundFA}}

\label{sec:lem2}

\begin{lemma}\label{lem:LowBoundFA} Fix the horizon $T$ and $\delta_{\mathrm{F}}$. Then, for any stopping rule $\tau$ that satisfies the first contraint in \eqref{eq:QCD}, $\forall\,T\in\mbb{N}$, $\forall\,\delta_{\mathrm{F}}\in\lp0,1\rp$, and $\forall\,d\in\lbp1,\dots,T\rbp$, there exists a $\tilnu\in\lbp0,\dots,T-d\rbp$ such that:
\begin{align}
\Pr_{\infty}\lp\tilnu\leq\tau< \tilnu+d\rp\leq\frac{\delta_{\mathrm{F}}}{\lfl T/d\rfl}.\label{eq:FA_window}
\end{align}
\end{lemma}

\begin{proof}
Suppose that $\exists\,d\in\lbp1,\dots,T\rbp$ such that $\forall\,\nu\in\lbp1,\dots,T-d\rbp$,
\begin{align}
\Pr_{\infty}\lp\nu\leq\tau<\nu+d\rp>\frac{\delta_{\mathrm{F}}}{\lfl T/d\rfl}.
\end{align}
Then
\begin{align}\nonumber
\Pr_{\infty}\lp\tau\leq T\rp&\geq\Pr_{\infty}\lp\tau< d\lfl T/d\rfl\rp\\\nonumber
&=\sum_{i=0}^{\lfl T/d\rfl-1}\Pr_{\infty}\lp di\leq\tau<d\lp i+1\rp\rp\\\nonumber
&>\sum_{i=0}^{\lfl T/d\rfl-1}\frac{\delta_{\mathrm{F}}}{\lfl T/d\rfl}\\
&=\delta_{\mathrm{F}}.
\end{align}
This leads to a contradiction since $\Pr_{\infty}\lp\tau\leq T\rp\leq\delta_{\mathrm{F}}$.
\end{proof}

\section{Proof of Lemma \ref{lem:sub_martin}}

\label{sec:lem3}

\vvv{This lemma is no longer needed.}

\begin{lemma}\label{lem:sub_martin}
$\forall\,\nu\in\mbb{N}$, under the probability measure $\Pr_{\nu}$, i.e., $X_{m}\diid f_{1}$ $\forall\,t>\nu$, then the sequence $\lp\prod_{i=\nu+1}^{\nu+m}\frac{f_{1}\lp X_{i}\rp}{f_{0}\lp X_{i}\rp}\rp_{m=1}^{\infty}$ is a nonnegative submartingale.
\end{lemma}

\begin{proof}
First, we show that $C\geq\textrm{kl}\lp f_{1};f_{0}\rp$, where $\textrm{kl}$ denotes the Kullback-Leiber (KL) divergence:
\begin{align}\nonumber
C&=\log\lp\E_{f_{1}}\lb\frac{f_{1}\lp X\rp}{f_{0}\lp X\rp}\rb\rp\\\nonumber
&\overset{(a)}{\geq}\E\lb\log\lp\frac{f_{1}\lp X\rp}{f_{0}\lp X\rp}\rp\rb\\
&=\textrm{kl}\lp f_{1};f_{0}\rp\label{eq:C_ineq}
\end{align}
where step $(a)$ is due to Jensen inequality.

Next, we can show that $\forall\,\nu,m\in\mbb{N}$,
\begin{align}\nonumber
    &\E_{\nu}\lb\prod_{i=\nu+1}^{\nu+m}\frac{f_{1}\lp X_{i}\rp}{f_{0}\lp X_{i}\rp}\bigg|X_{\nu+1},\dots,X_{\nu+m-1}\rb\\\nonumber
    &=\prod_{i=\nu+1}^{\nu+m-1}\frac{f_{1}\lp X_{i}\rp}{f_{0}\lp X_{i}\rp}\E_{\nu}\lb\frac{f_{1}\lp X_{\nu+m}\rp}{f_{0}\lp X_{\nu+m}\rp}\bigg|X_{\nu+1},\dots,X_{\nu+m-1}\rb\\\nonumber
    &=\prod_{i=\nu+1}^{\nu+m-1}\frac{f_{1}\lp X_{i}\rp}{f_{0}\lp X_{i}\rp}\E_{\nu}\lb\frac{f_{1}\lp X_{\nu+m}\rp}{f_{0}\lp X_{\nu+m}\rp}\rb\\\nonumber
    &=e^{C}\prod_{i=\nu+1}^{\nu+m-1}\frac{f_{1}\lp X_{i}\rp}{f_{0}\lp X_{i}\rp}\\\nonumber
    &\overset{(a)}{\leq} e^{0}\prod_{i=\nu+1}^{\nu+m-1}\frac{f_{1}\lp X_{i}\rp}{f_{0}\lp X_{i}\rp}\\
    &=\prod_{i=\nu+1}^{\nu+m-1}\frac{f_{1}\lp X_{i}\rp}{f_{0}\lp X_{i}\rp}\label{eq:sub_mart}
\end{align}
where step $(a)$ is due to \eqref{eq:C_ineq} and the fact that KL divergence is nonnegative. Since $\prod_{i=\nu+1}^{\nu+m}\frac{f_{1}\lp X_{i}\rp}{f_{0}\lp X_{i}\rp}$ is nonnegative $\forall\,\nu,m\in\mbb{N}$, \eqref{eq:sub_mart} proves Lemma \ref{lem:sub_martin}.
\end{proof}

\section{Proof of Lemma \ref{lem:d_inf}}

\label{sec:lem4}

\begin{lemma}\label{lem:d_inf}
    When $\delta_{\mathrm{F}},\delta_{\mathrm{D}}\in\lp0,1\rp$ satisfies $\delta_{\mathrm{F}}+\delta_{\mathrm{D}}<1$, $d^{*}\lp T,\delta_{\mathrm{F}},\delta_{\mathrm{D}}\rp\rightarrow\infty$ as $T\rightarrow\infty$
\end{lemma}

\begin{proof}
First, by the definition, $d^{*}\lp T,\delta_{\mathrm{F}},\delta_{\mathrm{D}}\rp$ is increasing with $T$ for fixed $\delta_{\mathrm{F}}$ and $\delta_{\mathrm{D}}$. Thus, either the limit of $d^{*}\lp T,\delta_{\mathrm{F}},\delta_{\mathrm{D}}\rp$ exists or it goes to infinity as $T\rightarrow\infty$.

We prove that $d^{*}\rightarrow\infty$ as $T\rightarrow\infty$ by contradiction. Assume that $d^{*}\lp T,\delta_{\mathrm{F}},\delta_{\mathrm{D}}\rp$ does not go to infinity as $T\rightarrow\infty$ when $\delta_{\mathrm{F}}+\delta_{\mathrm{D}}<1$. Then, for any $\bar{d}$ larger than the limit of $d^{*}$ as $T\rightarrow\infty$, for any $T\in\mbb{N}$, and for any $c>C$, by \eqref{eq:thm_1_sketch_4}, we have:
\begin{align}
\delta_{\mathrm{D}}&\geq1-\delta_{\mathrm{F}}-\frac{\bar{d}\exp\lp c\bar{d}\rp\delta_{\mathrm{F}}}{T-\bar{d}}-e^{-\bar{d}\lp c-C\rp}.\label{eq:lem_4_proof_1}
\end{align}
Then, by taking $T\rightarrow\infty$, we have:
\begin{align}
\delta_{\mathrm{D}}\geq1-\delta_{\mathrm{F}}-e^{-\bar{d}\lp c-C\rp}.\label{eq:lem_4_proof_2}
\end{align}
Last, by taking $\bar{d}\rightarrow\infty$, we have $\delta_{\mathrm{D}}\geq1-\delta_{\mathrm{F}}$. Since $\delta_{\mathrm{F}}+\delta_{\mathrm{D}}<1$, this assumption leads to a contradiction.
\end{proof}

\end{document}